\documentclass[useAMS,usenatbib]{mn2e}
\usepackage{times}
\usepackage{graphicx}

\def\aap{A\&A}%
\def\aj{AJ}%
\def\apj{ApJ}%
\def\apjl{ApJ}%
\def\apjs{ApJS}%
\def\apss{Astroph.~\& Space Sci.}

\def\iaucirc{IAU Circ.}%
\def\mnras{MNRAS}%
\def\pasp{PASP}%
\def\ssr{Space Sci.~Rev.}%
\def\zap{Zeitschr.~f\"ur Astroph.}%

\title[Life after eruption. IV.]{Life after eruption -- IV. 
Spectroscopy of 13 old novae}
\author[C. Tappert et al.]%
{
C. Tappert,$^{1}$\thanks{E-mail: claus.tappert@uv.cl}
N. Vogt,$^{1}$ 
M. Della Valle,$^{2}$ 
L. Schmidtobreick$^{3}$
and
A. Ederoclite$^{4}$
\footnotemark[1]\thanks{Based on observations with Euorpean Southern
Observatory (ESO) telescopes,
proposal numbers 083.D-0158(A), 089.D-0505(B), 090-D-0069(B)}\\
$^{1}$Instituto de F\'{\i}sica y Astronom\'{\i}a, Universidad de 
Valpara\'{\i}so, Avda. Gran Breta\~na 1111, 2360102 Valpara\'{\i}so, Chile\\
$^{2}$INAF - Osservatorio di Capodimonte, Salita Moiariello 16, 80131 Napoli,
Italy\\
$^{3}$European Southern Observatory, Alonso de Cordova 3107, 7630355 Santiago, 
Chile\\
$^{4}$Centro de Estudios de F\'{\i}sica del Cosmos de Arag\'on, Plaza San 
Juan 1, Planta 2, Teruel, E44001, Spain\\
}
\begin{document}

\date{Accepted. Received}

\pagerange{\pageref{firstpage}--\pageref{lastpage}} \pubyear{2013}

\maketitle

\label{firstpage}

\begin{abstract}
We present data on 13 post-nova systems. This includes the recovery 
via {\em UBVR} photometry of the five post-novae X Cir, V2104 Oph, V363 Sgr,
V928 Sgr and V1274 Sgr and their spectroscopic confirmation. We provide
accurate coordinates and finding charts for those objects. Additional
first-time or improved spectroscopic data are presented for V356 Aql, V500 Aql,
V604 Aql, V1370 Aql, MT Cen, V693 CrA, V697 Sco and MU Ser. Investigating
the behaviour of a few easily accessible parameters yields (limited)
information on the accretion state and the system inclination. We predict
that X Cir and V697 Sco are likely to reveal their orbital period via
time series photometry and that long-term photometric monitoring of V356 Aql,
V500 Aql, V1370 Aql and X Cir has a good chance of discovering outburst-like
behaviour in these systems.
\end{abstract}


\section{Introduction}
\defcitealias{tappertetal12-1}{Paper I}%
\defcitealias{tappertetal13-1}{Paper II}%
\defcitealias{tappertetal13-2}{Paper III}%

A classical nova eruption is a thermonuclear runaway on the surface of
the primary component of a cataclysmic variable (CV), i.e.~an interacting
white-dwarf/red-dwarf binary system. This is a bright event:
eruption amplitudes range between 8 and 14 mag, and the nova at maximum reaches
absolute magnitudes typically from $-7$ to $-9$ mag 
\citep{dellavalle+livio95-1}. Novae are thus usually well studied during 
maximum brightness and for a few weeks or months after. However, once the
brightness has declined to a stable 'quiescence' value \citep[which may %
be fainter, equal to, or brighter than the pre-eruption brightness;][]%
{collazzietal09-1}, detailed investigations become comparatively scarce, 
mostly because such now need a significant amount of time on large telescopes.
Because most Galactic novae are located in the Galactic disc and thus within
crowded fields, this has led to the situation that for a large 
fraction (currently about 50 per cent) of the $\sim$200 novae that were 
reported before 1980 the position is not known with sufficient accuracy 
to properly identify the post-nova.

However, the study of post-novae has the potential to provide the answer
to several questions concerning the influence of the nova eruption on the
CV hosting it \citep{sharaetal86-1,johnsonetal14-1} as well 
as concerning the physical parameters of the CV (magnetic field, masses, 
mass-transfer rate, chemical composition) affecting the recurrence time of the 
nova eruption \citep[e.g.][]{townsley+bildsten05-1}. This is especially
true since it is now generally accepted that the nova eruption represents
an integral part of CV evolution \citep[e.g.][]{pattersonetal13-2}. However,
a comparative study of the overall CV population with that of post-novae
even with respect to comparatively easily accessible parameters like the
orbital period \citep{townsley+bildsten05-1,tappertetal13-2} still lacks
statistical significance due to the small sample size of the latter systems.
We have thus set out to increase the sample of post-novae by identifying
candidates for the post-nova via photometric colour-colour diagrams and
confirming them spectroscopically. For more details on the motivation and
the methods see \citet[][hereafter Paper I]{tappertetal12-1}.

In order to minimize the contribution of the ejected material in the visual
wavelength range, we usually restrict our study to novae that were reported 
before 1980. However, in this paper, we additionally include spectroscopic 
data on a number of systems where the time between maximum brightness and
the spectroscopic observations amounts to less than 30 yr. In none of
those systems do we find signatures of the ejected material and they thus
equally fit our profile, in that they offer an undiluted view on the
underlying CV.

\section{Observations and reduction}

\begin{table*}
\begin{minipage}{2.0\columnwidth}
\caption[]{Log of observations.}
\label{obslog_tab}
\setlength{\tabcolsep}{0.15cm}
\begin{tabular}{@{}lllllll}
\hline\noalign{\smallskip}
Object & RA (2000.0) & Dec.~(2000.0) & Date & Filter/Grism 
& $t_\mathrm{exp}$ (s) & mag \\
\hline\noalign{\smallskip}
V356 Aql  & 19:17:13.71 & $+$01:43:21.7 & 1993 July 12     & B300 (1.5 arcsec)
& 2400             & --   \\
V500 Aql  & 19:52:28.00 & $+$08:28:46.2 & 1992 July 8      & B300 (2.0 arcsec)
& 1800             & --   \\
V604 Aql  & 19:02:06.39 & $-$04:26:43.7 & 2012 June 14     & 300V (LSS)
& 2 $\times$ 900   & 18.8R \\
V1370 Aql & 19:23:21.10 & $+$02:29:26.1 & 1993 July 12     & B300 (1.5 arcsec)
& 3000             & -- \\
MT Cen    & 11:44:00.24 & $-$60:33:35.7 & 2013 January 12  & 300V (LSS)
& 2 $\times$ 900   & 19.5R \\
X Cir     & 14:42:41.46 & $-$65:11:18.7 & 2009 May 20      & $U$/$B$/$V$/$R$ 
& 1800/900/300/180 & 19.3V \\
       &             &                  & 2013 March 13    & 300V (MOS)
& 2$\times$ 888    & 18.5R \\
       &             &                  & 2013 March 14    & 300V (MOS)
& 888              & 18.7R \\
V693 CrA  & 18:41:58.02 & $-$37:31:14.5 & 1993 July 11     & B300 (2.0 arcsec)
& 3600             & -- \\
V2104 Oph & 18:03:24.99 & $+$11:47:57.1 & 2009 May 22      & $U$/$B$/$V$/$R$ 
& 1800/900/300/180 & 20.9V \\
          &             &               & 2012 July 14     & 300V (LSS)
& 2$\times$ 2700   & 20.4R \\
V363 Sgr  & 19:11:18.90 & $-$29:50:37.8 & 2009 May 21+22   & $U$/$B$/$V$/$R$
& 3600/900/300/240 & 19.0V \\
          &             &               & 2012 April 29    & 300V (MOS)
& 2640             & 19.2R \\
V928 Sgr  & 18:18:58.09 & $-$28:06:34.9 & 2009 May 22      & $U$/$B$/$V$/$R$ 
& 1800/900/300/180 & 19.5V \\
          &             &               & 2012 May 17      & 300V (MOS)
& 2 $\times$ 2640  & 19.8R \\
V1274 Sgr & 17:48:56.19 & $-$17:51:51.0 & 2009 May 20      & $U$/$B$/$V$/$R$ 
& 1800/900/300/180 & 19.2V \\
          &             &               & 2012 June 21     & 300V (MOS)
& 2640             & 18.9R \\
V697 Sco  & 17:51:21.91 & $-$37:24:56.9 & 1992 July 8      & B300 (2.0 arcsec)
& 900              & -- \\
MU Ser    & 17:55:52.78 & $-$14:01:17.1 & 1993 July 11     & B300 (1.5 arcsec)
& 3600             & -- \\
\hline\noalign{\smallskip}
\end{tabular}
\end{minipage}
\end{table*}

The $U\!BV\!R$ photometric data were taken in 2009, May, using the ESO Faint 
Object Spectrograph and Camera \citep*[EFOSC2; ][]{eckertetal89-1} 
at the ESO-NTT, La Silla, Chile. Typically, a
series of $\ge$3 frames per filter was obtained. EFOSC2 suffers from a 
well-known central light concentration that makes proper flat-fielding a 
difficult and time-consuming process. Since our scientific goal requires only
moderately high photometric precision, we refrained from a flat-field 
correction, and only performed the subtraction of bias frames in the reduction 
process. Subsequently, the frames were corrected for the individual telescope
offsets and averaged using a 3$\sigma$ clipping algorithm to minimize the
effect of bad pixels. Photometric magnitudes for all stars in the fields on 
both the averaged data and the individual frames (to check for variability) 
were extracted using the aperture photometry routines in {\sc iraf}'s 
{\sc daophot} package and the standalone {\sc daomatch} and {\sc daomaster}
routines \citep{stetson92-1}. These magnitudes were calibrated using
observations of standard stars \citep{landolt83-1,landolt92-3}.

The spectroscopic data consist of two sets. The first one was taken on
two runs in 1992 and 1993 July, with EFOSC1 mounted on the ESO 3.6 m telescope
in La Silla, Chile. The B300 grating was used with a 1.5 arcsec and a
2.0 arcsec slit, resulting in a spectral resolution of $\sim$15 and $\sim$20
{\AA}, respectively, over a useful wavelength range of $\sim$3800--6900 {\AA}.
The efficiency curve of B300 peaks in the blue at about 4500 {\AA}. At that
time the EFOSC set-up included a Tektronix 512$\times$512 pixel CCD (\#26)
with a red efficiency curve peaking at $\sim$6900 {\AA}, resulting in an
approximately uniform efficiency over the whole wavelength range. Data 
reduction was performed with {\sc eso-midas} \citep{warmels92-1} and consisted 
of bias reduction and flat-field correction. The wavelength calibration that 
has been performed using a Helium-Argon lamp deviates somewhat blueward of
4100 {\AA}. The spectral energy distribution (SED) was corrected for the 
response function of the instrument and a formal correction for the 
atmospheric extinction was performed, but since the data were taken during 
non-photometric conditions no flux calibration was applied. For a few systems,
we find a number of weak and narrow emission lines that we attribute to the 
extraction process suffering from an imperfect subtraction of 
the night sky, leaving residuals in the 5000--6000 {\AA} range,
especially concerning the \mbox{[O\,{\sc i}]} $\lambda$5577.338 line.

The second set of spectroscopic observations was performed in service mode at 
the ESO-VLT using the FOcal Reducer/low dispersion Spectrograph 
\citep[FORS2;][]{appenzelleretal98-3} with grism 300V and a 1.0 
arcsec slit. Depending on the number of original candidates in the field, the 
instrument was operated either in long-slit (LSS) or in multi-object (MOS) 
mode. Reduction of the data consisted of subtraction of an averaged bias frame 
and division by a flat field that was previously normalized by fitting a cubic 
spline of high order to the response function. After optimal extraction 
\citep{horne86-1} with {\sc iraf}'s {\sc apall} routine, the spectra were 
wavelength calibrated using the spectra of comparison lamps. The typical 
wavelength range is $\sim$4100--9000 {\AA} at a spectral resolution of 
$\sim$11 {\AA}, measured as the full width at half-maximum (FWHM) of the 
calibration lines. Finally, the data were flux calibrated using observations
of spectrophotometric standard stars. Because the data were taken in service
mode, the red CCD mosaic (two 2k$\times$4k MIT CCDs) had to be used for the 
observations. As a consequence of its low efficiency, blueward of $\sim$4400, 
{\AA} the SED in that part is not reliable. Other than for the EFOSC data,
the acquisition frames (taken in the $R$ passband) were available for the FORS 
observations. This allowed for a photometric analysis and thus the
extraction of a differential $R$ magnitude that was calculated with respect
to a number of comparison stars. The apparent $R$ magnitude was subsequently
derived with respect to previous photometric data.

Table \ref{obslog_tab} summarizes the details of the observations sorted in
the order of the variable designation. The second and third column contain
the coordinates. For the novae with photometric observations (this includes
those with acquisition frames) they were determined performing astrometry
with Starlink's 
{\sc gaia}\footnote{\tt http://astro.dur.ac.uk/$\sim$pdraper/gaia/gaia.html} 
tool
(version 4.4.3) using the US Naval Observatory CCD Astrograph Catalog (UCAC)
version 3 \citep{zachariasetal10-1} and 4 \citep{zachariasetal13-1}. The 
typical rms resulted to $\le$0.24 arcsec. Coordinates for the
other systems were taken from the \citet{downesetal05-1} or
\citet{saitoetal13-2} catalogues. 
Column 4 states the date of the observations corresponding to the start of 
the night in local time. Columns 5 and 6 give the instrument 
configuration, i.e.~the filter and grism used, and the corresponding total 
exposure times, respectively. Finally, where available, the last columns 
states the brightness at the time of the observations with the letter at the 
end identifying the corresponding passband.

\section{Results}
\label{results_sec}

\begin{figure*}
\includegraphics[width=2.0\columnwidth]{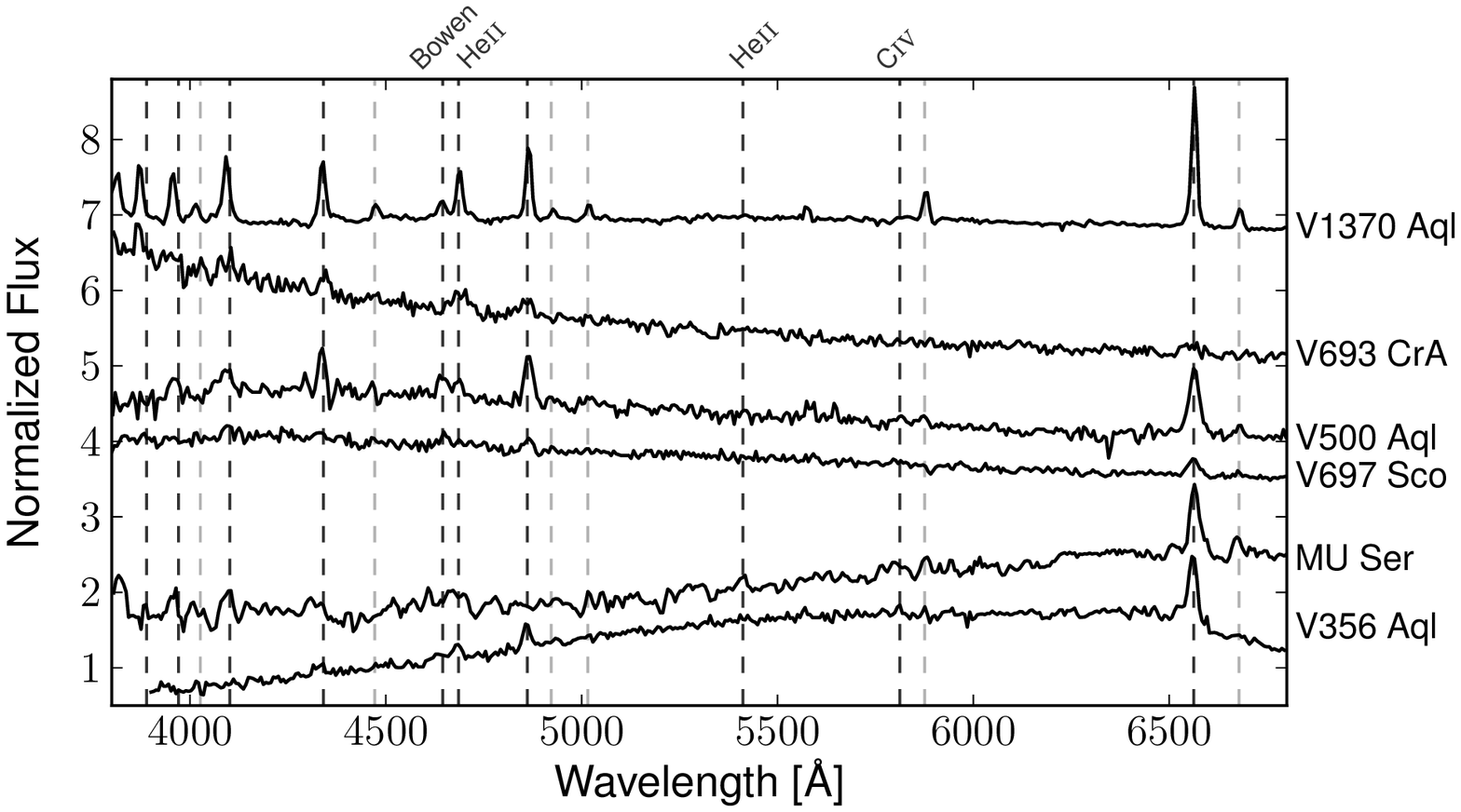}
\caption[]{EFOSC spectra. Unlabelled black vertical dashed lines mark the 
positions of Balmer emission, grey lines indicate the \mbox{He\,{\sc i}} 
series.}
\label{3p6sp_fig}
\end{figure*}

\begin{figure*}
\includegraphics[width=2.0\columnwidth]{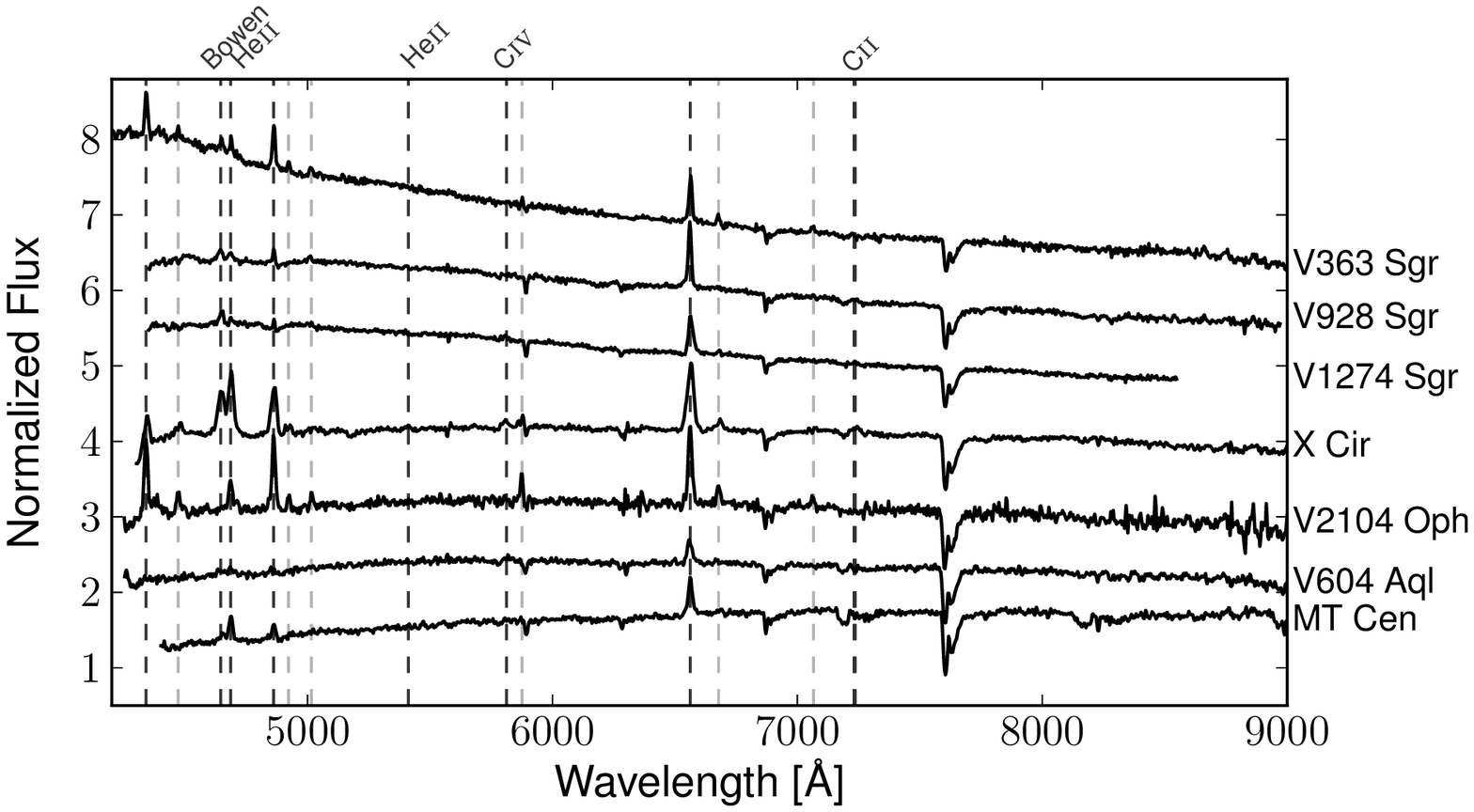}
\caption[]{FORS2 spectra. Unlabelled black vertical dashed lines mark the 
positions of Balmer emission, grey lines indicate the \mbox{He\,{\sc i}} 
series.}
\label{vltsp_fig}
\end{figure*}

\begin{table*}
\begin{minipage}{16cm}
\caption[]{Equivalent widths in angstroms of the principal emission lines.
A blank field means that the line was not included in the spectral range,
a '$-$' means that the line was not detected. Values in square brackets refer
to emission cores within absorption troughs.}
\label{eqw_tab}
\begin{tabular}{@{}lllllllllllll}
\hline\noalign{\smallskip}
Object    &  \multicolumn{3}{c}{Balmer} 
& \multicolumn{6}{c}{He\,{\sc i}} & Bowen/He\,{\sc ii} 
& He\,{\sc ii} & C\,{\sc iv}\\
          & 4340   & 4861     & 6563    & 4472   & 4922   & 5016   & 5876$^a$ 
& 6678 & 7065   & 4645/4686$^b$ & 5412   & 5812 \\
\hline\noalign{\smallskip}
V356 Aql  & 6(1)   & 7(1)     & 29(2)   & --     & 2(2)   & --     & --       
& 2(2)   &        & 12(3)       & --     & -- \\
V500 Aql  & 9(2)   & 18(2)    & 41(3)   & 1(1)   & 6(1)   & 5(1)   & 4(2)
& 4(2)   &      & 11(1)         & --     & 4(2) \\
V604 Aql  & --     & 1.5(3)   & 7.9(5)  & --     & --     & --     & 2.4(3)   
& 0.6(2) & --     & 5.2(6)      & --     & -- \\
V1370 Aql & 20(1)  & 20(1)    & 39.1(5) & 4.4(5) & 2.8(9) & 3.5(6) & 7.1(5)
& 5.4(6) &        & 23(1)       & --     & -- \\ 
MT Cen    &        & 4.9(9)   & 10.3(3) & --     & --     & --     & --       
& 1.0(3) & --     & 12.8(5)     & --     & -- \\
X Cir     & 11(2)  & 17.5(5)  & 28.1(4) & 3.5(5) & 3(1)   & 1.7(5) & 2.5(5)   
& 3.1(6) & 1.0(1) & 43.5(8)     & 0.8(3) & 2.6(3) \\
V693 CrA  & 5(1)   & 9(2)     & 11(3)   & 1.7(9) & --     & --     & --
& --     &        & 12(2)       & --     & -- \\
V2104 Oph & 18(1)  & 18(1)    & 18.7(7) & 6(1)   & 2.8(4) & 2.4(6) & 4.9(5)   
& 3.9(3) & 2.0(4) & 9.1(7)$^c$  & --     & -- \\
V363 Sgr  & 4.1(2) & 5.2(3)   & 11.5(5) & 0.9(2) & 0.8(2) & 0.8(2) & 0.9(1)   
& 2(0.5) & 2(0.5) & 3(0.5)      & --     & -- \\
V928 Sgr  &        & [1.8(1)] & 14.3(4) & --     & --     & 1.2(2) & --       
& 0.6(5) & --     & 4.5(4)      & --     & -- \\
V1274 Sgr &        & [0.4(1)] & 11.5(4) & --     & --     & 0.7(2) & --       
& 0.9(1) & --     & 4.4(4)      & --     & 0.9(2) \\
V697 Sco  & 2(2)   & [4(2)]   & 9.7(7)  & --     & 0.9(7) & --     & --
& 2.1(8) &        & 2.7(8)$^d$  & --     & -- \\
MU Ser    & 18(2)  & --       & 23(2)   & --     & --     & --     & --
& 4(1)   &        & 14(4)       & 5(2)   & 7(2) \\
\hline\noalign{\smallskip}
\end{tabular}
\\
$^a$ this line is mostly distorted by the adjacent Na\,{\sc I} absorption.\\
$^b$ the two components are not resolved.\\
$^c$ \mbox{He\,{\sc ii}} 4686 only; $^d$ Bowen blend only
\end{minipage}
\end{table*}

\begin{table}
\caption[]{Results of the $U\!BV\!R$ photometry.}
\label{ubvr_tab}
\begin{tabular}{@{}lllll}
\hline\noalign{\smallskip}
Object & $V$ & $U\!-\!B$ & $B\!-\!V$ & $V\!-\!R$ \\
\hline\noalign{\smallskip}
X Cir     & 19.29(12) & $-$0.45(02) & 0.48(03) & 0.45(04) \\
V2104 Oph & 20.94(08) & $-$0.57(03) & 0.50(02) & 0.41(03) \\
V363 Sgr  & 19.02(14) & $-$0.95(01) & 0.16(01) & 0.02(01) \\
V928 Sgr  & 19.51(08) & $-$0.55(01) & 0.37(01) & 0.22(02) \\
V1274 Sgr & 19.15(11) & $-$0.70(03) & 0.42(03) & 0.23(02) \\
\hline\noalign{\smallskip}
\end{tabular}
\end{table}

\begin{figure}
\includegraphics[width=\columnwidth]{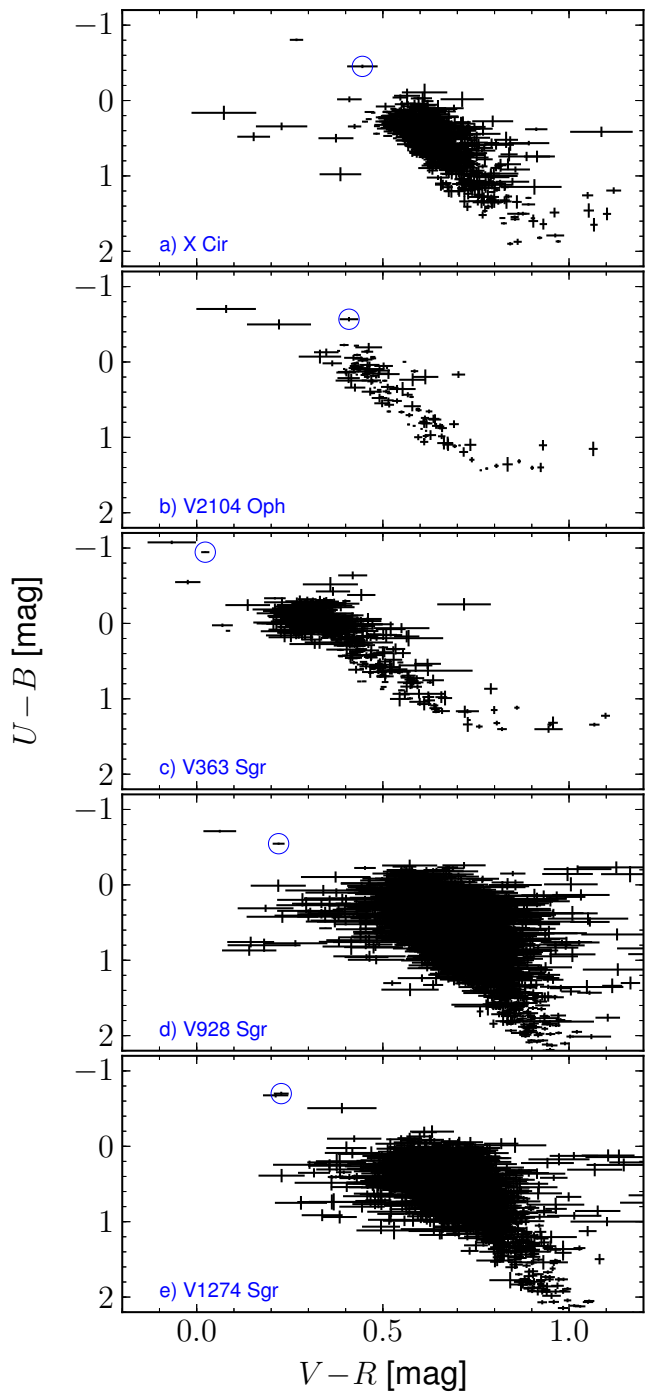}
\caption[]{Colour-colour diagrams. The confirmed post-novae are marked with
a circle.}
\label{ubvr_fig}
\end{figure}

In Figs.~\ref{3p6sp_fig} and \ref{vltsp_fig} we present the spectra taken with
the ESO 3.6 m telescope
and at the VLT, respectively. For systems with more than one spectrum,
the combined data are shown. Equivalent widths ($W_\lambda$) of principal
emission lines in these spectra are collected in Table \ref{eqw_tab}
\footnote{For convenience, we use positive values for $W_\lambda$. This is 
unambiguous since we exclusively refer to emission lines.}. The
corresponding errors were estimated with a Monte Carlo simulation by adding
random noise to the data and repeating the measurement a thousand times.
Fig.~\ref{ubvr_fig} presents the $U\!-\!B$/$V\!-\!R$ colour-colour diagrams
for the five systems with respective data, and Table \ref{ubvr_tab} details
the corresponding values.

For the analysis of the spectra, we have used the values for interstellar
extinction $E(B\!-\!V)$ from \citet{schlafly+finkbeiner11-1} using NASA's 
Infrared Science Archive web pages. The values are listed in Table 
\ref{prop_tab} and represent the average extinction within a 2$\times$2 degree
field. They were used to deredden the
spectra employing the corresponding {\sc iraf} task, which is based on the 
relations 
derived by \citet*{cardellietal89-1}. A standard value for the ratio of the
total to the selective extinction $R(V) = A(V)/E(B\!-\!V) = 3.1$ was
used. The dereddened spectra were fitted with a power law 
$F \propto \lambda^{-\alpha}$, restricting the continuum to wavelengths
5000--7200 {\AA} and masking strong emission and absorption lines.

We furthermore folded the flux calibrated (VLT) spectra with 
\citet{bessell90-1} passbands to obtain spectrophotometric magnitudes. A
comparison for those systems with information on photometric magnitudes
suggests slit losses $\le$0.5 mag.

\subsection{V356 Aquilae = Nova Aql 1936}
\label{v356aql_sec}

This relatively slow nova was discovered by N. Tamm, Kvistaberg Observatory, 
Denmark, on 1936, September 18 as an object of eighth magnitude 
\citep{stroemgren36-1,duerbeck87-1}. \citet{mclaughlin55-3} describes the
eruption light curve as a fast rise from July 17 to 19 followed by a more
gradual rise and irregular fluctuations over several months. The recorded
maximum value was 7.1 mag on October 3, although the primary maximum is
suspected to have occurred during an unobserved period between July 31 and
August 10. \citet{robinson75-1} lists the pre-eruption magnitude to 16.5 mag.
However, \citet{szkody94-2} finds the post-nova at $V$ = 18.3 mag in 1988,
and \citet*{ringwaldetal96-3} report $V$ = 17.9 mag in 1991. 
\citet*{woudtetal04-2} present high-speed photometry revealing flare-like 
outburst activity at time-scales of about 50 min, but no coherent 
periodicities. \citet{duerbeck+seitter87-1} describe the spectrum of V356 Aql 
as a blue continuum with strong H$\alpha$ and weaker H$\beta$ emission, 
as well as with a strong Bowen/\mbox{He\,{\sc ii}} emission component. 
\citet{ringwaldetal96-3} present a spectrum with medium strong H$\alpha$ 
($W_\lambda$ = 15 {\AA}). They also detect \mbox{He\,{\sc i}} $\lambda$6678, 
while the blue part has too low S/N to reveal any details.

Our spectrum taken at the ESO 3.6 m telescope shows emission lines over
a red continuum (Fig.~\ref{3p6sp_fig}). The latter differs strongly from the 
description of
\citet{duerbeck+seitter87-1} and the spectrum published by 
\citet{ringwaldetal96-3}. The spectrum was taken at a fairly high air mass
$M(z)$ = 2.1 and it is thus possible that the continuum slope is affected
by a non-parallactic position angle of the slit. With respect to the
emission lines, our data better fits that of \citet{duerbeck+seitter87-1},
with the H$\alpha$ emission being considerably stronger than in the
\citet{ringwaldetal96-3} data (Table \ref{eqw_tab}). In the blue part
of the spectrum we furthermore identify H$\beta$, the adjacent 
\mbox{He\,{\sc i}} $\lambda$4922 line, a medium strong 
Bowen/\mbox{He\,{\sc ii}} blend and H$\gamma$.
Higher Balmer lines fall victim to the strong Balmer decrement and the
low S/N. Apart from the usual residuals of the night sky lines in the 
5000--6000 {\AA} range, we also find a couple of emission spikes that 
reasonably 
coincides with the positions of \mbox{C\,{\sc iv}} $\lambda$5812 and 
\mbox{He\,{\sc i}} $\lambda$5876. Another, double, spike is close to 
\mbox{He\,{\sc ii}} $\lambda$5412. However, 
since in appearance (FWHM) and strength they are more similar to the 
night sky residuals than to the other intrinsic emission lines, we do not 
consider those spikes as sufficient evidence for the presence of those lines.

\subsection{V500 Aquilae (Nova Aql 1943)}
\label{v500aql_sec}

According to \citet{duerbeck87-1}, this object was detected by Hoffmeister on 
plates of the Sonneberg Observatory, Germany. The recorded maximum light of
6.55 mag was reached on 1943 May 2, although the real maximum can be suspected
to have occurred sometime before that. \citet{cohen85-1} detected a small shell 
around the ex-nova, with a diameter of the order of 2 arcsec, and an 
expansion velocity of 1380 km s$^{-1}$. \citet{szkody94-2} finds the post-nova 
at $V$ = 19.3 mag. Time-resolved photometry by \citet{haefner99-1} during two 
nights in August 1994 revealed that V500 Aql is an eclipsing binary with a 
period 
of 3.5 h and an eclipse depth of the order of 0.5 mag, with superimposed 
flickering. The only spectroscopic data to date come from the radial velocity 
study by \citet{haefner+fiedler07-1} that has a high spectral resolution of 
1.2 {\AA} but covers only a limited wavelength range of 4000-5000 {\AA}. The 
spectra show a weak Balmer series (H$\beta$ to H$\delta$) and a moderately
strong Bowen/\mbox{He\,{\sc ii}} blend. Complex changes at different orbital 
phases  
can be recognized, especially in the Balmer line profiles. In spite of being
an eclipsing system, the emission lines in almost all phase bins are
single peaked. This is similar to the behaviour of the SW Sex-type stars
\citep{thorstensenetal91-1} which represent the dominant population in the
3--4 h orbital period range \citep{rodriguez-giletal07-2}. The fact that the
period of V500 Aql lies within this regime thus fits well into this picture.

Our spectrum shows a blue continuum and unusually strong emission lines for
an old nova (Fig.~\ref{3p6sp_fig}). 
In contrast to the \citet{haefner+fiedler07-1} spectra, we can
clearly identify the \mbox{He\,{\sc i}} $\lambda\lambda$4471 and 4922 lines, 
which
perhaps are hidden in the lower S/N of their data. Like for V356 Aql 
(Section \ref{v356aql_sec}), we find a few emission spikes in the 5000--6000 
{\AA} range that we do not believe to be intrinsic emission lines due to their 
smaller FWHM. However, the components corresponding to the positions of
\mbox{C\,{\sc iv}} $\lambda$5812 and \mbox{He\,{\sc i}} $\lambda$5876 look 
sufficiently 
convincing. Their presence is also in agreement with the general appearance
of the spectrum, namely that a large number of \mbox{He\,{\sc i}} lines can be
identified, and that the Bowen C/N component is at least as strong as the
adjacent \mbox{He\,{\sc ii}} $\lambda$4686 line. 

\subsection{V604 Aquilae (Nova Aql 1905)}
\label{v604aql_sec}

This relatively fast nova was discovered on Harvard plates by 
W. Feming on 1905, August 31, but additional plates, taken a few 
weeks earlier, place the photographic maximum brightness of 8.2 mag to 
mid-August 1905, fading rapidly afterwards. \citet{duerbeck87-1} gives the 
references on the early literature. 
Photometric $BV\!R\!J\!K$ data of the object were obtained by 
\citet{szkody94-2}, determining the minimum light to $V \sim$ 19.6 mag.
A 3.5 h CCD light curve was published by \citet{haefner04-1}. It exhibits 
erratic short-term variations with amplitudes up to about 0.25 mag, but also a 
longer lasting structure that bears some resemblance to an orbital hump.
 
The spectrum of the star marked in \citet{downesetal05-1} shows weak and narrow
emission lines (Fig.~\ref{vltsp_fig}). 
The continuum peaks at $\sim$5600 {\AA}. However, the 
extinction amounts to $E(B\!-\!V) \sim$0.7 mag, and after dereddening
the spectrum with this value, we find a slope that corresponds to a power-law
index $\alpha \sim$2.1 which is a rather typical value for high $\dot{M}$
post-novae \citepalias{tappertetal12-1}. The slope flattens slightly towards
the blue part of the spectrum. For the potentially magnetic nova V630 Sgr such 
has been interpreted as a signature for a disrupted inner accretion disc 
\citep{schmidtobreicketal05-2}, but in V604 Aql the difference in slopes
is much less clear.

Comparison of the acquisition frame with the UCAC4 catalogue indicates an 
$R$-band brightness $\sim$18.8 mag. Folding the spectrum with Bessell 
passbands yields $V$ = 19.8 mag and $R$ = 19.3 mag. Taking into account a
probable slit loss $\le$0.5 mag as observed for the other FORS2 spectra, we 
find that 
these results agree well with each other. Those values are not significantly
different from those measured by \citet{szkody94-2} in 1988 April ($V$ = 19.6 
and $R$ = 19.1 mag) especially considering the amplitude of the variability 
observed by \citet{haefner04-1}.

\subsection{V1370 Aquilae (Nova Aql 1982)}
\label{v1370aql_sec}

As reported in \citet{kosaietal82-2}, the nova was discovered by M.~Honda in 
Japan, 1982 January 27. An analysis of the eruption light curve by 
\citet*{rosinoetal83-1} suggests that maximum light was reached a few days
earlier and up to two magnitudes brighter than the recorded value of 6.5 mag.
They classify the object as a fast nova with $t_3 \sim$10 d. For the 
corresponding literature during the first years after outburst maximum see 
\citet{duerbeck87-1} and \citet{snijdersetal87-1}. A comprehensive analysis
on the eruption of V1370 Aql can be found in \citet{smits91-1}, combining
several aspects including the light curve, the high expansion velocity,
as well the spectral development in the  UV, IR and optical range, and the 
elemental abundances in the ejected shell. \citet*{stropeetal10-1} file the
nova into the 'D' class, meaning that there was a dip in the eruption light
curve, indicating the formation of dust. \citet{szkody94-2} finds the
post-nova at $V$ = 18.0 mag seven years after maximum brightness, while 
\citet{ringwaldetal96-3} lists $V$ = 18.5 mag two years later. The latter
authors furthermore present a spectrum that shows very strong Balmer emission
lines ($W_\lambda$ = 43 {\AA} for H$\alpha$) and the presence of the 
\mbox{He\,{\sc i}} 
series, but -- only nine years after the eruption -- already no sign of 
nebular emission. 

Our spectrum in Fig.~\ref{3p6sp_fig}
was taken one year after the one presented in 
\citet{ringwaldetal96-3}. It is quite unusual for an old nova, in that it
shows an extremely hydrogen and \mbox{He\,{\sc i}} rich emission line forest. 
While
in other old novae, the higher Balmer lines $>$H$\delta$ tend to be hidden
in the absorption lines of an optically thick disc, the Balmer series in
V1370 Aql can be followed up to the limit of our blue spectral range, the 
last line that can be clearly identified being H11 $\lambda$3771. The Balmer 
decrement in general is also comparatively flat, indicating an optically thin 
disc. The only exception represents H$\alpha$ which has about twice the 
strength of H$\beta$, suggesting the presence of an additional H$\alpha$ 
emitter. The \mbox{He\,{\sc i}} series is remarkably well pronounced. Even 
weak lines
like $\lambda$4026, $\lambda$4388 (as an emission hump on the red wing of
H$\gamma$) and $\lambda$5048 are clearly discernable. Overall, V1370 Aql
has more the appearance of a (comparatively) low mass-transfer CV than
of an old nova, with the exception of the presence of a quite prominent 
Bowen/\mbox{He\,{\sc ii}} component and of the absence of other low $\dot{M}$ 
indicators
like \mbox{Ca\,{\sc ii}} and \mbox{Fe\,{\sc i}} emission 
\citep[see e.g.~the spectrum of %
GZ Cnc which otherwise has similar characteristics as V1370 Aql; ][]%
{tappert+bianchini03-1}. The signatures of an optically thin disc are
the more remarkable as the emission lines are quite narrow, indicating that
the system inclination is low, and thus that a large fraction of the disc 
should be able to contribute to the radiation from the system. An optically
thin and thus intrinsically faint disc would also support above idea that
the system at maximum was considerably brighter than the recorded 6.5 mag,
since the resulting eruption amplitude of 11.8 mag (Table \ref{prop_tab})
appears too low for such a system and thus should be regarded as a lower limit.

\subsection{MT Centauri (Nova Cen 1931)}
\label{mtcen_sec}

The spectroscopic confirmation of this nova was already reported in
\citetalias{tappertetal12-1} and we refer the reader to that publication for
a summary on the available data. In the meantime, the object has also been
identified in the VISTA Variables in the V\'{\i}a L\'actea Survey (VVV),
with \citet{saitoetal13-2} providing infrared magnitudes. The spectrum in
\citetalias{tappertetal12-1} suffered from very low S/N and thus revealed very
little detail on the emission lines. The FORS spectrum presented in 
Fig.~\ref{vltsp_fig} is of much higher quality and allows for the additional
identification of the \mbox{He\,{\sc i}} $\lambda$6678 emission line. We 
furthermore
note that the Bowen/\mbox{He\,{\sc ii}} blend is clearly dominated by the 
helium emission.

As in \citetalias{tappertetal12-1}, the spectrum shows a red slope that is
entirely due to interstellar reddening. After the corresponding correction,
the SED is that of a high $\dot{M}$ system with a power-law index of
$\alpha$ = 3.44(1). This is somewhat less than what was determined for the
\citetalias{tappertetal12-1} spectrum ($\alpha$ = 4.45(7)). However, the major
portion of this difference is due to the use of the revised extinction value 
from \citet{schlafly+finkbeiner11-1} in this paper with respect to 
\citet*{schlegeletal98-2} that was used in \citetalias{tappertetal12-1},
i.e.~$E(B\!-\!V)$ = 1.38(2) instead of 1.61(2). Applying the old value for
the correction of the new spectrum yields $\alpha$ = 4.12(1).

Even with the new extinction value, the dereddened slope is much steeper than
for the other systems. However, other indicators of high $\dot{M}$, like the
weakness of the Balmer emission, point to a somewhat lower $\dot{M}$ than
e.g.~for V928 Sgr and V1274 Sgr (Sections \ref{v928sgr_sec} and 
\ref{v1274sgr_sec}, respectively), both with a significantly lower slope
(Table \ref{prop_tab}). It appears thus reasonable to assume that only
part of the dust that causes the high extinction in this region lies in
the line of sight to MT Cen, and that applying the full extinction value
overcorrects the reddening of its SED.

\subsection{X Circinis (Nova Cir 1926)}
\label{xcir_sec}

\begin{figure}
\includegraphics[width=\columnwidth]{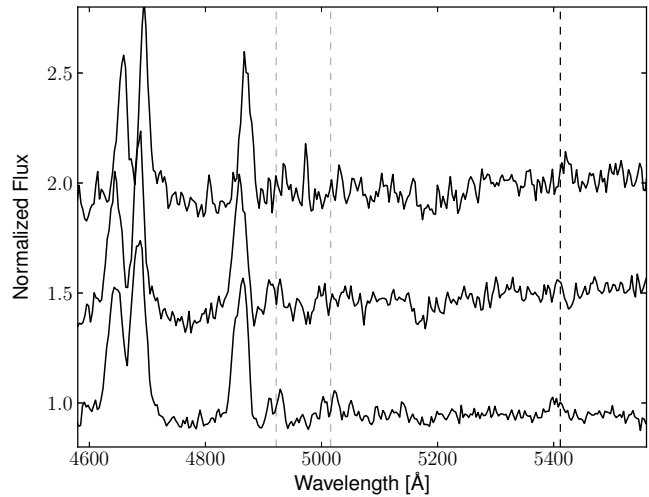}
\caption[]{Close-up on the blue part of the three spectra of X Cir. 
\mbox{He\,{\sc i}} positions are marked by dashed grey lines, the position of 
\mbox{He\,{\sc ii}} 5412
is indicated by the black dashed line.}
\label{xcirsp_fig}
\end{figure}

The nova was discovered spectroscopically on an objective prism plate taken at 
La Paz, Bolivia, 1927, May 21, by \citet{becker29-3}. At the time of 
discovery, it had a photographic brightness of about 11.0 mag. A search on
archive plates by \citet{cannon30-4} revealed that the maximum brightness
of 6.5 mag was reached on 1926 September 3. \citet{duerbeck87-1} lists
it thus as a slow nova with $t_3$ = 170 d. \citet{woudt+warner02-5} report
a candidate for the post-nova, based on flickering activity. However,
a spectrum taken by \citet{mason+howell03-1} of this object presents only
absorption lines and a comparatively red continuum peaking at 
$\sim$6000 {\AA}, making this an unlikely candidate for the post-nova.

Our $U\!BV\!R$ photometry of X Cir yields a number of potential candidates
(Fig.~\ref{ubvr_fig}, top). While the object that turned out to be the nova
is certainly one of the two most interesting ones colour-wise, we had it
nevertheless listed as one of the minor candidates, because first, with
$\sim$1.3 arcmin it lies at quite a distance from the reported coordinates, 
and secondly, the point spread function (PSF) on all photometric frames and 
also in the
2-D spectrum appears slightly elongated at an angle of roughly 50$^\circ$
as counted from north to east. Unfortunately, the PSFs of the images in
different filters vary considerably probably due to insufficient focus
correction, and thus do not allow for an analysis of the extended object
in this respect. On the other hand, the 2-D spectrum suggests that the
emission lines are confined to the eastern part of the PSF. Our preliminary
conclusion is that this is a close visual binary with a separation $<$0.6
arcsec, with the nova being the north-eastern component.

The spectrum of X Cir is quite remarkable (Fig.~\ref{vltsp_fig}). For a nova, 
it presents 
comparatively strong hydrogen emission lines that allows even for the
detection of the blue end of the Paschen series (from Pa13 $\lambda$8665 on).
The contribution of \mbox{He\,{\sc ii}} is considerable, as is evident from the
presence of the $\lambda$5412 line. We furthermore detect carbon in the form
of the \mbox{C\,{\sc iv}} $\lambda$5812 line and the 
\mbox{C\,{\sc ii}} $\lambda\lambda$7231/7236
doublet. This combination makes for a particularly strong 
Bowen/\mbox{He\,{\sc ii}}
blend, similar to what is observed in the old nova V840 Oph 
\citep{schmidtobreicketal03-5}.

In Fig.~\ref{xcirsp_fig} we have plotted a close-up on the blue part of the 
three individual spectra. The first spectrum taken (on HJD 245\,6365.70523)
is shown at the bottom, the middle spectrum was obtained 3.0 h later
(HJD 245\,6365.83078), and the top spectrum was taken 27.6 h after the first 
one (HJD 245\,6366.85626). The spectra have been normalized by dividing 
through the mean value of the $\lambda\lambda$4500--5600 {\AA} range and 
displaced vertically. We note that the \mbox{He\,{\sc i}} lines are 
distinctively 
double peaked with the central valley in some lines reaching the continuum 
level.
All other lines are single peaked. In the case of the Balmer series, however,
it can be clearly seen that there is an additional emission component
superposed on the (potentially) double-peaked profile. The same component is
also present in the \mbox{He\,{\sc i}} lines. Note how in the bottom spectrum 
the
red peak is broader than the blue one, while the opposite can be observed
in the middle spectrum. The asymmetry in the H$\beta$ line behaves in the
same manner. In old novae, an irradiated secondary star is a likely source of
such additional emission, but with the limited present data other
possibilities can of course not be excluded. The \mbox{He\,{\sc ii}} and 
carbon lines,
on the other hand, while single-peaked, do not present any discernable 
additional contribution but a broad single-peaked profile. 

We furthermore observe that the \mbox{He\,{\sc ii}} $\lambda$5412
line is well above continuum level in the bottom spectrum, below an
unambiguous detection in the middle spectrum and again clearly present in the
top spectrum. Because the individual exposure times amount to only 15 min,
this could represent an orbital effect. This becomes even more likely when
one considers that the double-peaked \mbox{He\,{\sc i}} lines let suspect a
comparatively high system inclination. Comparing the relative brightness 
of the continuum between $\lambda\lambda$5000 and 5400 {\AA,} we find that the
middle and top spectra are fainter than the first by $\sim$0.4 and $\sim$0.7
mag, respectively. From the acquisition frames, we compute the brightness in
the $R$ band to 18.47(02) mag just before the first spectrum was taken and
to 18.71(02) mag just before the third spectrum. Such differences are not
unusual for CVs seen at high inclination, and the brightness is thus
also consistent with the photometry from 2009 ($R$ = 18.84(12) mag; see
Table \ref{ubvr_tab}).

\subsection{V693 Coronae Austrinae (Nova CrA 1981)}
\label{v693cra_sec}

As reported by \citet{kozaietal81-2} this fast nova was discovered by 
M. Honda in Japan, 1981 April 2 at a visual brightness of 7.0, while it was 
invisible one day earlier. \citet{williamsetal85-1} observed its eruption
over several months with the International Ultraviolet Explorer (IUE) satellite
and on the basis of the derived
elemental abundances conclude that the nova contains a massive ONeMg white 
dwarf. A re-analysis of archival IUE data by \citet*{vanlandinghametal97-1}
agrees with this result as does the recent work by \citet{downenetal13-1},
who use actual thermonuclear eruption models in combination with elemental 
abundance determinations during the eruption to determine the white dwarf
mass to $\le$1.3 M$\odot$. \citet{schmidtobreicketal02-1} report
$V$ = 21.0 mag for the post-nova.

Due to the faintness of the object, our spectrum taken at the ESO 3.6 m
telescope is of limited quality (Fig.~\ref{3p6sp_fig}). 
It shows a steep blue continuum with
rather weak Balmer lines. Of the \mbox{He\,{\sc i}} series, only the line at
4472 {\AA} can be reasonably well identified. The emission lines are
very broad, which could indicate a high inclination, or, since the
FWHMs for different lines span quite a range (26 {\AA} for H$\gamma$, 
32 {\AA} for H$\beta$, 43 {\AA} for H$\alpha$), possibly contamination
from the shell material. Overall, this is certainly a high $\dot{M}$ CV.

\subsection{V2104 Ophiuchi (Nova Oph 1976)}
\label{v2104oph_sec}

The nova was discovered on 1976 September 23 at a visual brightness
of 8.8 mag \citep{osawa+kuwano76-1}. In spite of being a comparatively
young nova, the post-eruption coverage is poor. \citet{huth76-1} measures
a pre-maximum photographic brightness of 9.3 mag for 1976 September 22, and a 
fading to 12.4 mag for October 21. This yields an upper limit for $t_3$ of
28 d. \citet{robertsonetal00-1} report a candidate for the post-nova
with $B$ = 21.0 mag that is H$\alpha$ bright according to a private 
communication by Downes. A search for a shell by \citet{downes+duerbeck00-1}
remained unsuccessful, but they report the same object at $V$ = 20.5.

The colour-colour diagram (Fig.~\ref{ubvr_fig}) indicates a couple of blue,
but very faint, objects, while the \citet{robertsonetal00-1} candidate is 
comparatively red and about half a magnitude fainter than previously reported
(Table \ref{ubvr_tab}). Nevertheless, its spectrum shows the typical emission 
lines that confirm the nova (Fig.~\ref{vltsp_fig}).  Like in X Cir
(Section \ref{xcir_sec}), the Balmer 
and \mbox{He\,{\sc i}} lines are comparatively strong, but in the case of 
V2104 Oph they 
are much more narrow. The typical FWHM of a line amounts
to $\sim$16~{\AA} and lies thus only slightly above the spectral resolution.
We furthermore note that there is no detectable Bowen component, but that
\mbox{He\,{\sc ii}} $\lambda$4686 is clearly present, if not particularly 
strong. The
continuum is comparatively flat, which appears to be intrinsic since there is
little interstellar extinction in this region of the sky 
(Table \ref{prop_tab}), and does not show any detectable contribution from 
the secondary star.

The strengths of the lines and the flat continuum suggest that V2104 Oph has
comparatively low $\dot{M}$. We also note that the (dereddened) spectral slope 
cannot be fitted by a single power law, but shows two vastly different slopes
blue- and redward of $\sim$5670 {\AA}, with both individual power indices
being well below a steady-state disc (Table \ref{prop_tab} and Section 
\ref{disc_sec}). The narrowness of the emission lines translates
to a narrow range of sampled velocities. Since we do not see any correlation
of the FWHM with the line species or degree of ionization, this indicates
that the object is seen at a low inclination. This makes it unlikely that
the brightness difference of our data with respect to the measurements by
\citet{robertsonetal00-1} and \citet{downes+duerbeck00-1} is due to
orbital variability.

\subsection{V363 Sagittarii (Nova Sgr 1927)}
\label{v363sgr_sec}

An objective prism spectrum taken on 1927 September 30 showed emission lines in 
an object with a brightness of $\sim$11 mag \citep{becker30-3}. An archival 
search by \citet{walton30-1} confirmed the suspicion of a nova event and 
revealed that the photographic maximum brightness of 8.8 mag had already been 
reached on August 3 and perhaps even somewhat earlier (because of a lack of 
observations from July 1 to August 2). \citet{duerbeck87-1} gives two potential
candidates for the post-nova.

However, V363 Sgr was recovered via its position in the colour-colour diagram 
(Fig.~\ref{ubvr_fig}) roughly
40 arcsec outside the area marked in \citet{downesetal05-1} and thus 
corresponding to neither of the two \citet{duerbeck87-1} candidates. The 
spectrum (Fig.~\ref{vltsp_fig}) shows a steep blue continuum and weak narrow
emission lines. Apart from the Balmer and \mbox{He\,{\sc i}} series, we can 
identify
the Bowen blend and \mbox{He\,{\sc ii}} $\lambda$4686. Contrary to the other
systems in this paper, these two are almost fully separated in V363 Sgr.
The spectral appearance suggests that V363 Sgr is seen at a similar
low inclination as V2104 Oph, but accretes at significantly higher rates.

\subsection{V928 Sagittarii (Nova Sgr 1947)}
\label{v928sgr_sec}

\citet{burwell47-1} reported this slow nova as an H$\alpha$ bright star of
nineth magnitude based on an object prism plate taken on 1947 May 16. 
\citet{bertaud47-1} measures
a visual brightness of $\sim$8.9 mag on May 24. \citet{campos+sanchez87-1}
provide finding charts of variable stars in the constellation of Sagittarius,
one of which (their object 'L') \citet{morel87-2} identifies with V928 Sgr.
He provides revised coordinates that match the third possible position given
by \citet{duerbeck87-1}.
  
The colour-colour diagram of the field indicates a couple of blue objects,
with the nova turning out to be the brighter of the two roughly 3 arcsec
west and 29 arcsec south of above position. The spectrum
in Fig.~\ref{vltsp_fig} shows a comparatively steep blue continuum that
after dereddening gains a slope that within the errors is identical to
that of V363 Sgr (Table \ref{prop_tab}). However, while that system shows a
rich emission line spectrum, very few lines can be identified in V928 Sgr.
This is in part due to the low blue efficiency of the CCD affecting the
spectrum already blueward of 4400 {\AA} so that of the Balmer series only
H$\alpha$ and H$\beta$ lie within the useful spectral range. There, we
can observe a very strong Balmer decrement, with the H$\beta$ emission lying
partly within an absorption trough. We furthermore identify the 
Bowen/\mbox{He\,{\sc ii}}
blend. Other than in most other systems where the \mbox{He\,{\sc ii}} line is 
the 
stronger of the two \citep{ringwaldetal96-3}, the corresponding equivalent 
widths in V928 Sgr are $W_\lambda$ = 1.8 and 1.2 {\AA} for the individual 
peaks of the Bowen blend and \mbox{He\,{\sc ii}}, respectively. The 
\mbox{He\,{\sc i}} series is
also peculiar, in that the line at 5016 {\AA} is by far the strongest one, with 
in fact the line at 6678 {\AA} being the only other, barely, discernable 
\mbox{He\,{\sc i}} emission within our spectral range.

\subsection{V1274 Sagittarii (Nova Sgr 1954 No.2)}
\label{v1274sgr_sec}

\citet{wild54-1} discovered this nova as a 10.5 mag star on 1954 August 30. 
\citet{wenzel92-4} identified the nova on Sonneberg patrol plates, and by
combining them with previous data found that while the object was at
12 mag on August 2 and 3, five observations over a range of 38 d, from 
August 26 to October 3, always showed it at roughly 10.4 mag. This would
suggest a classification as a slow nova.
 
There are two blue stars with very similar colours in the field of V1274 Sgr
(Fig.~\ref{ubvr_fig}). The nova is the one that is slightly bluer in $U\!-\!B$
and slightly redder in $V\!-\!R$. It is also closer to the originally suspected
position, being just inside the circle from the \citet{downesetal05-1} chart.
The spectrum of the other object shows a blue continuum with Balmer and
(probably interstellar) \mbox{Na\,{\sc I}} absorption. With an
FWHM $\sim$25
{\AA}, the Balmer lines appear too narrow for a white dwarf, and thus this is 
probably an early B star.

The nova shows the appearance of a high $\dot{M}$ system, with few and
weak emission lines and H$\beta$ being visible only as a narrow emission
core within an absorption trough (Fig.~\ref{vltsp_fig}). 
The blue continuum becomes significantly
steeper after dereddening due to the high interstellar extinction in this
region (Table \ref{prop_tab}). We furthermore note the presence of the same
carbon emission lines as in X Cir (\mbox{C\,{\sc iv}} $\lambda$5812 and 
\mbox{C\,{\sc ii}} 
$\lambda\lambda$7231/7236) but also that \mbox{He\,{\sc ii}} is much weaker 
than in
that system. As a consequence, the Bowen component is much stronger than the
adjacent \mbox{He\,{\sc ii}} $\lambda$4686 line ($W_\lambda$ for the individual
peaks are 2.9 and 1.3 {\AA}, respectively).

\subsection{V697 Scorpii (Nova Sco 1941)}
\label{v697sco_sec}

The nova was discovered by \citet{mayall46-2} on Harvard objective prism plates
from 1941 March 9 as an object of 10.2 mag. Analysis of the light curve and
the spectra suggest that the maximum eruption brightness occurred sometime
before that date. \citet{duerbeck87-1} estimates the true maximum brightness 
to $\sim$8 mag and classifies it as a very fast nova with $t_3 <$15 d. 
\citet{warner+woudt02-5} obtained high speed photometry during a total of six
nights in May and June 2001. They found the ex-nova at brightnesses
$V$ = 19.5--20.1 mag, about 3 mag fainter than previously reported. Their 
light curves are characterized by rather strong variability with amplitudes up 
to 0.5 mag. A Fourier analysis of these observations reveals two 
periodicities, simultaneously present. Based on these observations, the 
authors consider V697 Sco as an intermediate polar with an orbital period 
$P_\mathrm{orb}$ = 4.49 h and a spin period of the white dwarf 
$P_\mathrm{rot}$ = 3.31 h. Such proximity of $P_\mathrm{rot}$ and 
$P_\mathrm{orb}$ is observed only in a small number of CVs, e.g.~also in the 
system EX Hya \citep*{vogtetal80-1}, while most other intermediate polars have 
$P_\mathrm{rot} \sim\,0.1~P_\mathrm{orb}$ 
\citep[][update 7.20, 2013]{ritter+kolb03-1}. 
\citet{saitoetal13-2} give revised 
coordinates and {\em JHK} photometry values from the VVV survey. 

Our spectrum (Fig.~\ref{3p6sp_fig}) shows a moderately blue continuum with 
weak emission lines.
Unfortunately the low S/N does not reveal much detail. However, we do 
find that the H$\beta$ emission line is embedded in shallow absorption 
troughs. This, and the relative strength of the Bowen blend indicates a high 
$\dot{M}$. We furthermore remark on the apparent absence of 
\mbox{He\,{\sc ii}} 
$\lambda$4686 emission. Considering the suspected high $\dot{M}$ and magnetic 
nature of the system, this comes as somewhat of a surprise. On the other hand, 
the above-mentioned intermediate polar EX Hya inhabits only a very weak 
\mbox{He\,{\sc ii}} $\lambda$4686 emission line itself 
\citep[e.g.][]{williams83-1}, 
and such a weak contribution in V697 Sco might be hidden in the noise. 
Clearly, further high S/N data are much desirable for this system.

\subsection{MU Serpentis (Nova Ser 1983)}

This extremely fast nova was discovered by \citet{wakuda83-1} on 1983 February 
22 at a brightness of 7.7 mag. Since it was fainter then 11 mag the day 
before, this 
value will be close to maximum brightness. \citet*{schlegeletal85-2} analyse 
the eruption light curve to find an upper limit for the maximum to $\sim$7.3 
mag and derive $t_3 =$ 5.3(1.6) d. To our knowledge, there are no further
data on the post-nova.

In Fig.~\ref{3p6sp_fig}, we present a spectrum that was taken 10 yr after 
the eruption.  It shows a red continuum
slope that we largely attribute to the high extinction in that region
(Table \ref{prop_tab}). The spectrum has fairly low S/N and we have thus
smoothed it with a 3 pixel box filter prior to the analysis. We detect 
moderately strong H$\alpha$, H$\gamma$ and even higher Balmer lines, but to 
our astonishment, we do not find any line at the position of H$\beta$. 
The higher S/N at the red end of the spectrum also reveals \mbox{He\,{\sc i}} 
$\lambda$6678, while the bluer lines of that series are probably hidden in
the noise. We furthermore report the likely presence of \mbox{He\,{\sc ii}} 
$\lambda$5412 and possibly also \mbox{C\,{\sc iv}} $\lambda$5812. Thus, MU Ser 
has the 
appearance of a high $\dot{M}$ post-nova. However, like for V697 Sco (Section 
\ref{v697sco_sec}), higher S/N data will be needed to properly
analyse the spectroscopic characteristics.

\section{Discussion}
\label{disc_sec}

\begin{table*}
\begin{minipage}{2.0\columnwidth}
\caption[]{Selected parameters of the post-novae. See Section \ref{disc_sec}
for details.}
\label{prop_tab}
\begin{tabular}{@{}llllllllllll}
\hline\noalign{\smallskip}
Object & $m_\mathrm{max}$$^a$ & $m_\mathrm{min}$ & $\Delta m$ & $\Delta t$ 
& $t_3$ & $t_2$ & $E(B\!-\!V)$ & $\alpha$ & FWHM(H$\alpha$) 
& $M_\mathrm{max}$ & $M_\mathrm{min}$ \\
 & (mag) & (mag) & (mag) & (yr) & (d) & (d) & (mag) & & ({\AA}) & (mag) 
& (mag)\\
\hline\noalign{\smallskip}
V356 Aql  & 7.1p     & 18.0$V$ & 10.9     & 57  & 140      & 36      
& 0.40(2)   & --                        & 25 & $-$7.2    & 3.7 \\
V500 Aql  & 6.6p     & 19.3$V$ & 12.7     & 49  & 42       & 19      
& 0.147(1)  & --                        & 30 & $-$8.0    & 4.7\\
V604 Aql  & 8.2p     & 19.6$V$ & 11.4     & 107 & 25       & 8       
& 0.69(1)   & 1.62(4) / 2.29(2) $^b$    & 23 & $-$8.7    & 2.7\\
V1370 Aql & $<$6.5v  & 18.3$V$ & $>$11.8  & 11  & $\sim$10 & 7       
& 0.41(2)   & --                        & 16 & $-$8.8    & $<$3.0\\
MT Cen    & 8.4p     & 19.8$V$ & 11.4     & 82  & $\sim$10 & 11      
& 1.38(2)   & 3.44(1)                   & 20 & $-$8.6    & 2.8\\
X Cir     & 6.5p     & 19.3$V$ & 12.8     & 87  & 170      & 160     
&  0.42(2)  & 1.52(1)                   & 31 & $-$6.8    & 6.0\\
V693 CrA  & 7.0v     & 21.0$V$ & 14.0     & 12  & 18       & 6       
& 0.0985(6) & --                        & 43 & $-$8.8    & 5.2\\
V2104 Oph & 8.8v     & 20.9$V$ & 12.1     & 36  & $<$29    & 6       
&  0.15(1)  & $-$1.69(5) / 0.80(3) $^b$ & 19 & $-$8.8    & 3.3\\
V363 Sgr  & 8.8p     & 19.0$V$ & 10.2     & 85  & 64       & 22      
& 0.103(4)  & 2.43(1)                   & 15 & $-$7.8    & 2.4\\
V928 Sgr  & 8.9p     & 19.5$V$ & 10.6     & 65  & 150      & 88      
&  0.39(1)  & 2.440(9)                  & 15 & $-$6.9    & 3.7\\
V1274 Sgr & 10.4p    & 19.2$V$ & 8.8      & 58  & $\gg$38  & $\gg$20 
&  0.52(5)  & 2.88(1)                   & 22 & $>$$-$7.9 & $>$0.9\\
V697 Sco  & $\sim$8p & 20.0$V$ & $\sim$12 & 51  & $<$15    & 8       
& 0.43(2)   & --                        & 34 & $-$8.8    & 3.2\\
MU Ser    & 7.7v     & --      & --       & 10  & 5        & 2.5     
& 0.77(4)   & --                        & 27 & $-$8.8    & --\\
\hline\noalign{\smallskip}
\end{tabular}
\\
$^a$ $p$: photographic, $V$: $V$-band, $b$: blue, $v$: visual\\
$^b$ first value for the blue part, second for the red part (see the text)
\end{minipage}
\end{table*}

\begin{figure}
\includegraphics[width=\columnwidth]{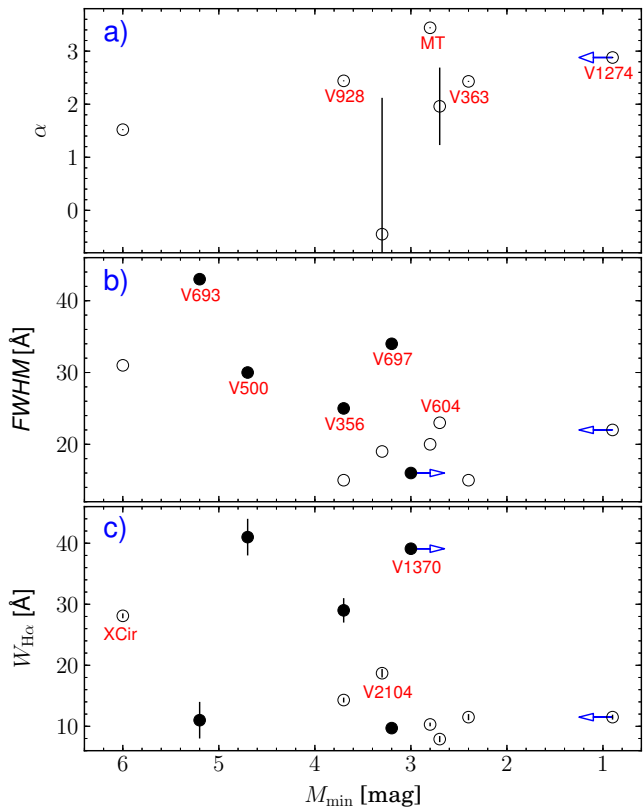}
\caption[]{The absolute minimum magnitude $M_\mathrm{min}$ versus the slope 
parameter $\alpha$ of the reddening corrected continuum (top), versus the 
FWHM of the H$\alpha$ emission line (middle), and versus its 
equivalent width (bottom). The systems of our sample with spectrophotometric
calibration are marked by open circles, those without by solid circles.}
\label{minmag_fig}
\end{figure}

In Table \ref{prop_tab}, we have selected several properties of the novae.
Column 2 gives the recorded maximum brightness as taken from the references
specified in the corresponding chapters above. Column 3 gives the brightness
of the post-nova, either from our own photometric data, or as the average
of the data found in the literature. $\Delta t$ in column 4 represents the
time from the eruption to the date of the spectroscopic data, while $t_3$
gives the time of decline to 3 mag below maximum brightness as listed in
\citet{duerbeck87-1} or \citet{stropeetal10-1}. The time of decline to 2 mag
in column 6 was determined from the eruption light curves, but for V928 Sgr and
V1274 Sgr, where $t_2$ was calculated from $t_3$ using the equations given
in \citet{capacciolietal90-5}. Details on the 
parameters in columns 7 and 8 are given in the introduction to Section 
\ref{results_sec}. Column 9 contains the FWHM of the H$\alpha$ emission 
line that can serve as an indicator for the system inclination. In contrast to
\citetalias{tappertetal12-1}, we have not used H$\beta$ here, because in a
few of the here present systems this line is found partly in absorption.
Since both decline time and the (absolute) maximum brightness $M$ of the nova 
eruption can be assumed to mostly depend on the mass of the
white dwarf \citep{livio92-8}, it is possible to find a relation
between those two parameters, the famous maximum-magnitude versus 
rate-of-decline relation. We have used the corresponding equation 
derived by \citet{dellavalle+livio95-1} to calculate the absolute maximum 
magnitudes $M_\mathrm{max}$ in column 10. Adding the eruption amplitude 
$\Delta m$ finally yields the absolute minimum magnitudes $M_\mathrm{min}$ 
in column 11. We do not attempt to derive distances because the proper 
correction for the interstellar absorption is unknown. The example of MT Cen 
(Section \ref{mtcen_sec}) shows that this needs a careful analysis of 
interstellar absorption lines and thus high-resolution and high-S/N data. 

While a proper study of possible relations between the parameters in
Table \ref{prop_tab} needs a sample of more significant size than
currently available, in Fig.~\ref{minmag_fig} we use them as a check
on the consistency of our results reported in Section \ref{results_sec}.
For example, the exponent $\alpha$ of the power law fitted to the continuum 
slope should reflect the brightness of the accretion disc (if present), and 
should thus be related to $M_\mathrm{min}$ because the brightness in the 
visual range of a CV is dominated by the luminosity of the accretion disc. 
The top plot of Fig.~\ref{minmag_fig} indeed shows a tendency that brighter 
systems have a steeper continuum slope. The two systems with large error bars 
for $\alpha$ are V604 Aql and V2104 Oph, where two different slopes were needed 
to fit the continuum (Sections \ref{v604aql_sec} and \ref{v2104oph_sec}). For 
the plot, we have used the average value, with the error bars corresponding to
the difference between the values. The arrows in the plot mark systems
with upper or lower limits in the parameters used to derive $M_\mathrm{min}$.

In the middle plot of Fig.~\ref{minmag_fig}, we test as a second example the 
behaviour of the FWHM of the H$\alpha$ emission line as a function of 
$M_\mathrm{min}$. The faintness of a post-nova can either be due to the 
absence of an accretion disc (i.e.~in polars), due to the disc being 
intrinsically faint indicating a low $\dot{M}$, or due the system being seen at 
high inclination (or, of course, a combination of those). Because a high 
system inclination $i$ yields a larger range of projected values $v \sin i$ 
for both the velocities in the disc and the velocities of the stellar 
components, the FWHM of the emission lines can be used as an indicator 
for $i$. In the corresponding plot, we find that it is likely that the 
faintness of V693 CrA (which has the highest FWHM) is at least in part 
due to a high $i$ and that it does not reflect a low $\dot{M}$. We will
come back to this in the next paragraph. The object with the faintest 
$M_\mathrm{min}$, X Cir, on the other hand, is also seen at high inclination. 
However, the difference in brightness to the other systems appears too large 
to be an exclusive effect of a high $i$ and it 
is thus likely that the system additionally inhabits an intrinsically faint 
disc. Last, not least, V697 Sco has comparatively broad emission lines, but 
still is placed among the brighter systems in our sample. We can thus conclude 
that this post-nova still drives a very high $\dot{M}$. 

Finally, the equivalent width $W_\lambda$ of the emission lines has been
proven a fairly good indicator for $\dot{M}$ in disc systems, with 
low-$\dot{M}$ presenting strong emission lines and high-$\dot{M}$
correspondingly weak ones \citep{patterson84-1}. In the bottom plot of 
Fig.~\ref{minmag_fig}, we show the respective values of the H$\alpha$ emission 
(Table \ref{eqw_tab}). We find that most of the systems in our samples appear 
to be high $\dot{M}$ systems with weak emission lines ($W_\lambda < 
20~\mathrm{\AA}$). Only four post-novae (X Cir, V500 Aql, V356 Aql and V1370 
Aql) can be suspected to have comparatively low $\dot{M}$.

Plotting $\alpha$, FWHM and $W_\lambda$ versus a common parameter also
allows us to easily extract information on their respective relations. For
example, we find that $\alpha$ and $W_\lambda$ show the expected roughly inverse
relationship (with the exception of the extreme 'hybrid' $\alpha$ system 
V2104 Oph). We furthermore find our above suspicion confirmed that in spite of 
its faintness V693 CrA still inhabits an intrinsically bright accretion disc,
as evidenced by its weak emission lines, and that the former is mainly due to 
the system seen at high $i$, as indicated by then line's width. On the
other hand, V1370 Aql appears to be a low $\dot{M}$ system seen approximately
face-on, as is suggested by its faint $M_\mathrm{min}$, low FWHM and
strong $W_\lambda$.

\section{Summary}

We have presented the recovery of five post-nova CVs with previously uncertain
or unknown identifications. The candidates have been selected via {\em UBVR}
photometry and confirmed spectroscopically. We furthermore have included
first or improved spectroscopic data for another eight objects. With this
we have further increased the number of confirmed post-novae. However, about
2/3 of the roughly 150 southern novae that were reported before 1980 still
lack spectroscopic confirmation or even an identification of candidates for
the post-nova, and there is thus still a long way to go until the post-novae
make up a sample of statistical significance.

While the establishing of such a sample represents the main purpose of our
project, we additionally find a number of systems that show peculiar 
properties that make them deserving of further, more detailed, investigation. 
Perhaps the most remarkable system in the present collection is X Cir. The
profiles of the emission lines suggest that the hydrogen and 
\mbox{He\,{\sc i}} lines
form in the outer, cooler, parts of the accretion disc, with an additional
component affecting the line profiles more (H) or less (\mbox{He\,{\sc i}}) 
from
possibly an irradiated secondary or a bright spot. On the other hand, the
\mbox{He\,{\sc ii}} and carbon lines will originate in a more confined 
high-temperature
region that does not allow for double peaks. Such could be either the innermost
parts of the accretion disc, or, if X Cir should turn out to be an intermediate
polar, the impact region of the gas stream on the surface of the white dwarf.
The distinctively double-peaked \mbox{\,He{\sc i}} lines indicate a 
comparatively
high system inclination. This suggests that the orbital period should be
accessible by the means of time series photometry. On the other hand,
high-resolved spectroscopic time series would be desirable to clarify the 
origin of the additional emission components, but such endeavour will be 
difficult due to the faintness of the object.

We have derived a number of parameters and investigated possible relations 
between them in order to gain information especially on the accretion state 
and the system inclination. Most of the systems included here appear
to drive a high mass-transfer rate and/or are seen at comparatively low
inclinations. However, X Cir and  V693 CrA have broad emission lines that 
suggest a higher inclination. Their orbital period could thus be comparatively 
easily accessible via time series photometry, such as has been already
performed successfully for the two novae with similar broad emission lines,
V500 Cyg and V697 Sco \citep[][respectively]{haefner99-1,warner+woudt02-5}.
Furthermore, V356 Aql, V500 Aql, V1370 Aql and X Cir show comparatively strong
emission lines and flat continuum slopes that indicates a lower mass-transfer
rate than for other post-novae. Those systems could harbour instable
accretion discs that allow for outburst-like behaviour 
\citep*{honeycuttetal95-1,honeycuttetal98-3,tappertetal13-1} and thus may be 
worth of long-term photometric monitoring.

\section*{Acknowledgements}
Thanks to Brian Skiff for bringing to our attention a couple of mistakes 
contained in the first arXiv upload.

We are indebted to the ESO astronomers who performed the service observations
of the VLT data.

Many thanks to Antonio Bianchini and Elena Mason for interesting discussions
especially concerning X Cir.

This research was supported by FONDECYT Regular grant 1120338 (CT and NV).
AE acknowledges support by the Spanish Plan Nacional de Astrononom\'{\i}a y 
Astrof\'{\i}sica under grant AYA2011-29517-C03-01. 

We gratefully acknowledge ample use of the SIMBAD data base, 
operated at CDS, Strasbourg, France, and of NASA's Astrophysics Data System 
Bibliographic Services. {\sc iraf} is distributed by the National Optical 
Astronomy Observatories. We thank the providers and maintainers of
OpenSuSE and Ubuntu Linux operating systems.


\appendix

\section{Finding charts}

We present the finding charts for the novae with previously ambiguous or
unknown positions. The images correspond to the photometric $R$-band data from
Table \ref{ubvr_tab}, but for V693 CrA, where the $V$-band data was used that
had been analysed in \citet{schmidtobreicketal02-1}. A finding chart for
MT Cen can be found in \citetalias{tappertetal12-1}. The other systems can be 
unambiguously identified from the charts presented by \citet{downesetal05-1}.
The nova MU Ser represents the only exception. While its position is marked
in the finding chart of \citet{duerbeck87-1}, the object itself is not
visible on that chart. The one by \citet{downesetal05-1} does not identify
the nova either, but only marks a certain area. Since the acquisition frames
of our spectroscopic data are lost, it falls to future photometric studies
to provide a proper finding chart. 

\begin{figure*}
\includegraphics[width=1.8\columnwidth]{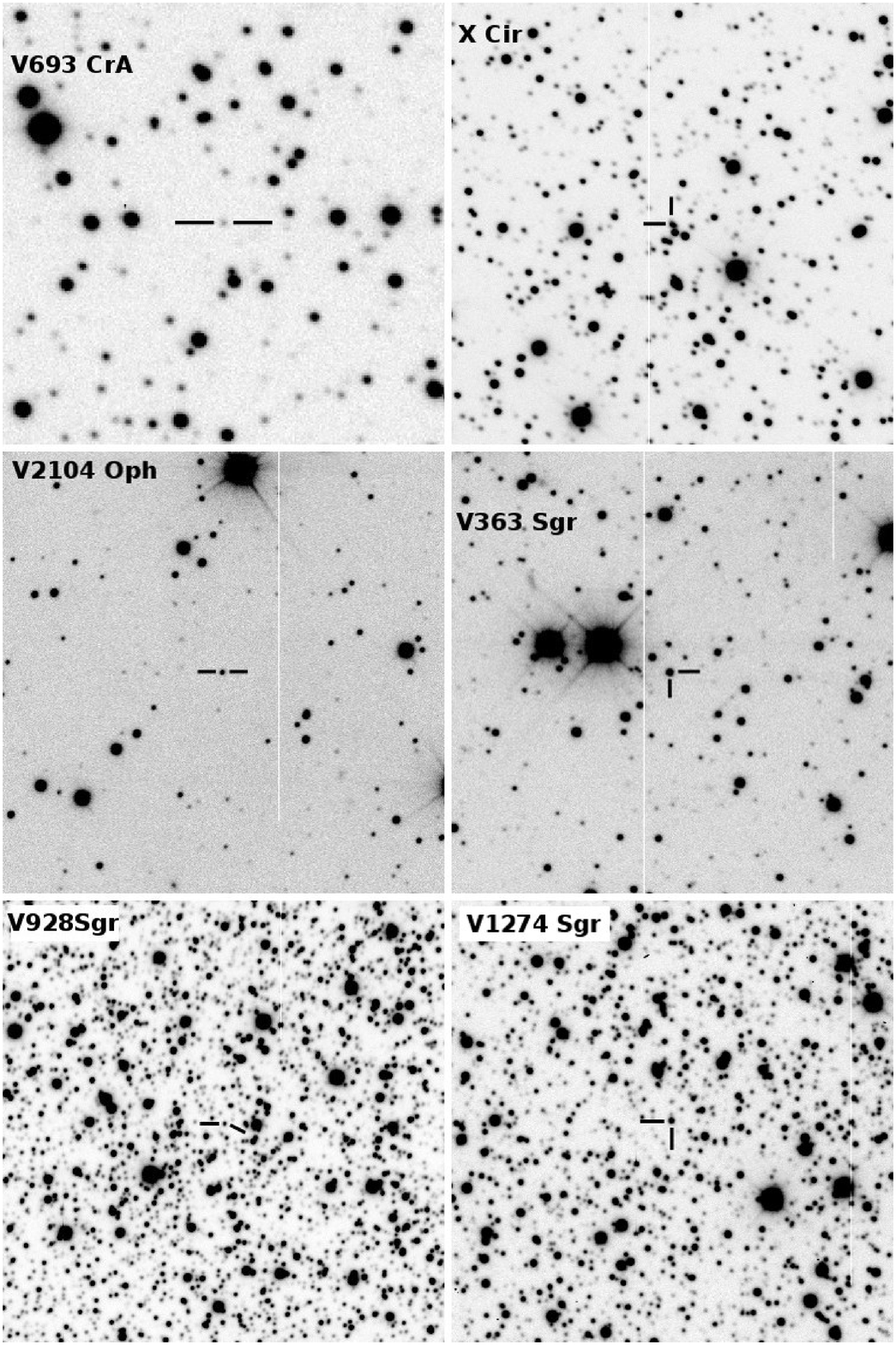}
\caption[]{Finding charts for the recovered old novae as labelled. The size of 
a chart is 1.5 $\times$ 1.5 arcmin$^2$, and the orientation is such that north 
is up and east is to the left.}
\label{fcall_fig}
\end{figure*}

\label{lastpage}

\end{document}